\documentclass{JINST}

\usepackage{graphicx}

\title{Study of radioactivity built-up and decay with singles time-stamped data}

\author{  Sangeeta Das$^a$, Arkajyoti De$^a$\thanks{Presently at  Indian Institute of Technology Delhi, New Delhi -110016,  INDIA}, 
Balaram Dey$^a$, Sathi Sharma$^a$, Anik Adhikari$^b$, S. S. Alam$^c$, Arkabrata Gupta$^b$,
Y. Sapkota$^b$, Ananya Das$^b$, A. Saha$^c$\thanks{Presently at  Indian Institute of Technology Bombay, Mumbai -400076,  INDIA}, Dibyadyuti Pramanik$^d$, 
T. Bhattacharjee$^c$, Abhijit Bisoi$^b$, S. Sarkar$^b$,  M. Saha
Sarkar$^a$\thanks{Corresponding author.}\\
\llap{$^a$}Saha  Institute  of  Nuclear  Physics, HBNI,
Kolkata - 700064, INDIA\\
\llap{$^b$}Indian Institute of Engineering Science and Technology, Howrah -711103, INDIA\\
\llap{$^c$}Variable Energy Cyclotron Centre, HBNI, Kolkata - 700064, INDIA\\
\llap{$^d$}Haldia Institute of Technology, Haldia - 721657, INDIA\\

  E-mail: \email{maitrayee.sahasarkar@saha.ac.in}}

\abstract{Time-stamped  data  have been used to estimate the decay half-lives of  radioactive $^{118m}$Sb and $^{64,66,68}$Ga nuclei. These nuclei are populated through the reaction $^4$He (E$_{lab}$ = 32 MeV) + $^{nat}$In and p (E$_{lab}$ = 10 MeV) + $^{nat}$Zn, respectively. The $\gamma$-rays emitted by the excited daughter nuclei are detected by a single high purity germanium (HPGe) detector and the data are acquired using a CAEN 5780M desktop digitizer. Half-lives of different orders of magnitudes for the above mentioned nuclei are obtained from both the decay of parent  and the growth of daughter nuclei. It is observed that the half-lives obtained from both these techniques match fairly  well and also are consistent with the reported values. It is demonstrated that the time-stamped data can be useful  while studying certain half-lives which are too long for measurement using in-beam electronic technique and too short for decay studies in off-beam methods. The data  also have a special relevance for constant monitoring of beam current and beam-tuning.}


\begin{document}
\section{Introduction}

The  decay half-life of a radioactive nucleus is one of the fundamental observable in nuclear spectroscopy.  
Lifetimes of radioactive nuclei are related to transition probabilities which give a quantitative estimation of the overlap of two wavefunctions (parent and daughter states) 
being connected by a transition ($\beta$-decay, say) operator. In several  astrophysical  processes, $\beta$-decay of exotic nuclei plays decisive role in the path of nucleosynthesis. Although study of $\beta$-decay half-lives has been pursued since long, there are still uncertainties in experimental results, even for nuclei close to the line of stability. For reliable  theoretical predictions, especially for exotic nuclei,  precise experimental data  in this field is most important.  These transition probabilities also provide  a wealth of information about the parent and daughter nuclei such as their internal structure, shape, deformation.  Reproduction of level lifetimes (electromagnetic transition probabilities) and $\beta$-decay half-lives (decay due to weak interaction), thus  in turn, provide crucial tests for validation of various theoretical models. Therefore, the precise measurement of lifetime is of key importance. 

 Recently, the impressive evolution in digital electronics, particularly the availability of high precision flash Analog to Digital Converters (ADC) and Field Programmable Gate Array (FPGA) has stimulated the usage of digital pulse processing (DPP) method in nuclear spectroscopy at a significant rate \cite{li-halflife,ne-halflife}.

In general, to measure short-lived activities,  targets are irradiated for  specific periods of time compatible to estimated half-lives. Decay data are then acquired in appropriate time steps to determine the half-lives. However, for a situation where a set of nuclei are populated having unknown half-lives, spanning a range from minutes to days, it is sometimes difficult to choose a time step for counting.  In the present work,  the half-lives of  different nuclei ($^{118m}$Sb, $^{64}$Ga, $^{66}$Ga and $^{68}$Ga) with half-lives ranging from few minutes to several hours, have been determined from   singles time-stamped data. Data from two different experiments have been used to demonstrate the effectiveness of the present technique.  The data have been accumulated in an  off-line experiment after irradiation and also continuously from in-beam to off-beam condition during an in-beam experiment. Later by analyzing list mode singles data with time-stamp, we could determine the half-lives of different orders and also could clearly differentiate the in-beam and off-beam situations. 

The measured half-lives have been compared with earlier results \cite{nndc} obtained using standard techniques with analog electronics in order to test the applicability of the present technique.

\section{Experimental details}
Two different experiments were performed to acquire the necessary data. Both of them have been done at Variable Energy Cyclotron Centre (VECC), Kolkata. One of them involved decay spectroscopic study of 
long-lived isotopes populated through irradiation. The other experiment was an in-beam study. 

\subsection{Experimental Setup: Detector and Data acquisition }
An HPGe detector  has been used to measure the decaying $\gamma$-rays from the excited target. Experimental data have been acquired using a CAEN 5780M \cite{CAEN} desktop digitizer (14 bit, 16k channel and 100 MS/s). The desktop module houses two high voltage power supplies of opposite polarities ($\pm$5 kV, 300 $\mu$A) and two  preamplifier  power supplies ($\pm$12 V, $\pm$24 V).  It is well equipped with DPP-PHA (Digital Pulse Processing-Pulse Height Analyzer) firmware that enables it to provide not only precise energy and timing information but also a portion of waveform and the other traces for fine-tuning of PHA settings \cite{CAEN}.

The exchange of data and control command between the digitizer and personal computer was executed by connecting an USB cable. The digitizer can also be connected via a CAEN proprietary optical bus. Via USB, the data can be transferred up to a rate of 30 MB/s, whereas optical link offers Daisy-chain capability with a transfer rate up to 80 MB/s \cite{CAEN}. As in our case, the digitizer is connected only with a single detector, using USB cable is the simplest choice. 

The output signal of the detector is directly fed into real-time digital pulse processing system (DPP), which is based on high speed digitizer, digital filter and analog to digital (A/D)  
converters (ADC). In contrast to conventional system (where A/D conversion takes place subsequent to analog processing), in these DPP units, A/D conversion is performed as early as possible without affecting the quality of the signal. In the present digitizer, the signals are continuously sampled at  100 MHz rate and the pulse heights are  saved  with 14 bit precision.
The readout or  digital processing is mainly handled by dedicated algorithms implemented in FPGAs, such as digital filters, pulse shapers, constant fraction discriminators, analog to digital converters etc. These algorithms are used to extract different necessary parameters like amplitude or the arrival time of the detected pulse from large data streams.
Whenever the pulse height crosses the set threshold value, a trigger is generated.  Pulse shaping procedure starts using a trapezoid filter \cite{trapezoidal}, that is applied on an input signal to transform the typical exponential pulse produced by charge sensitive preamplifier into trapezoidal shape. Basically, the pulse height difference between the flat top of trapezoid and signal base line is proportional to the amplitude of input pulse containing information of energy released by the incident radiation ($\gamma$ in the present case) in the detector. The role played by this filter is analogous to traditional Shaping Amplifier. The Pulse Height Analysis (PHA) algorithm includes  online baseline restoration, ballistic effect correction and manage the pile-up of online data acquisition \cite{CAEN}.

To determine the arrival time of the pulse, use of constant fraction discriminator (CFD) in the conventional system is replaced by an algorithm that calculates the second derivative of the input signal. The purpose of the derivation is to subtract the baseline part to avoid low frequency fluctuation. The zero crossing of the pulse is independent of pulse amplitude \cite{CAEN}. The zero crossing of the second derivative of input signal provides  accurate  arrival-time with an uncertainty equivalent to one unit of sampling time \cite{Mutti}. Whenever the pulse crosses the base line, a trigger signal is generated, producing a time-stamp.
The digital time-stamped data are saved with 10 ns time resolution. 

\subsection{Methodology}

The half-life of  the parent nucleus (an isomeric state in case of $^{118}$Sb) is generally determined from its decay which follows the equation 
\begin{equation}
\label{eqn1}
 N_m^t = N_m^0 exp(-\lambda_m t),
\end{equation}
where N$_m^0$ is the number of parent nuclei present at time t = 0  and $\lambda_m$ is decay constant which represents the rate at which the parents transmute to the daughter nuclei. 

The above equation can be also written as 
\begin{equation}
\label{eqn2}
 ln (N_m^t) = ln (N_m^0) -\lambda_m t,
\end{equation}

For a stable daughter (or if the decay half-life of daughter is much much larger than that of parent), one can also generate the growth curve of daughter nuclei from the data instead of following the decay of the parent nuclei. In this process, the number of daughter nuclei produced will  increase exponentially and reach a saturation value following the relation

\begin{equation}
\label{eqn3}
 N_d^t= N_m^0 [1 - \exp(-\lambda_m t)], 
\end{equation}

where N$_{d}^t$ is the number of daughter nuclei produced in time t. The half-life can be obtained either from the decay of parent nuclei or growth of daughter nuclei. The variation of count rate as a function of time is considered in case of the decay of parent nucleus, while the growth of the daughter nucleus is obtained by accumulating the data by increasing the time-bin of digitizer time-stamped data.  In the present study, we have followed both the decay of parent nucleus and the growth of daughter nucleus.
 
To follow decay, the count rates at different instants of time are determined from experimental data. Let, $T_o$ be the starting time of measurement at which the count rate is first measured. 
Usually spectra for small intervals of time  ($\Delta T$) starting from the initial time of counting  are generated ($T_o$ to $T_o+\Delta T$, $T_o+\Delta T$ to $T_o+2 \Delta T$) and areas under 
the most intense $\gamma$-peak in these decay spectra  are plotted as a function of time.

However, in the present work, somewhat different technique has been adopted for convenience in analysis of time- stamped data. From $T_o$ till the end of counting, same time interval  ($\Delta T$) 
has not been considered for all instants. Instead of time,  number of total events in the decay spectra are kept constant and thus the time bins at different instants varied. This means that at the beginning of the decay, the rate  is faster and smaller time bins are needed to  have 3 million (3M say,) decay events. On the other hand as time proceeds, same number of events result from longer time bins. 
Finally, count rates (number of decay events/ time of accumulation)  for each of these bins  are calculated and plotted as function of time to get the half-life.

The earlier equation (Eqn. \ref{eqn1}) thus reduces to,
\begin{eqnarray}
\label{eqn4}
 \frac{dN_m^t}{dt} = -\lambda_m N_m^0\exp(-\lambda_m t) 
\end{eqnarray}
The decay constant 
or half-life is determined by fitting the curve with the exponential function  (Eqn. \ref{eqn4}) by using $\chi^2$-minimization to ensure best fit.

For growth, spectra for equal number of decay events corresponding to different  intervals of $\Delta T$ are added consecutively starting from the initial time of counting ({\it i.e.} $T_o$ to $T_o+\Delta T$, $T_o$ to $T_o + \Delta T$ +$\Delta T^\prime$  ...)  and a set of growth spectra are generated. 
Thus, as time proceeds, the area of the $\gamma$-peak of interest increases i.e. statistics improves and  error decreases.

The time-stamped data recorded using the digital data acquisition system have been used to determine the nuclear lifetime of the other nuclei having short half-lives (short enough to measure after stopping the beam and long enough for electronic timing measurement) produced in an in-beam reactions. 
 Artificial radioactivity can be produced through  a nuclear reaction, by bombarding highly energetic beam (such as proton, neutron, alpha particles or heavy ions) on a stable nucleus (used as a target). 
The stable target nuclei ($N_t$) are  transformed into radioactive nuclei ($N_m$) that decay to another stable nucleus ($N_d$). In general, the accumulation of radioactivity during the irradiation process can be expressed as the difference between the rates of  production of  the radioactive nucleus  and  its spontaneous radioactive decay. The rate of production of a particular residual nucleus ($N_m$) is proportional to the flux of beam ($\Phi$), the activation cross-section ($\sigma_m$) and the number of 
target atoms (N$_{t}$). These residual radioactive nuclei also start decaying as soon as they are produced. In an in-beam experiment, 
 data have been accumulated continuously from in-beam to off-beam condition. After the beam stopped,  the decay rate of the short-lived isotopes have 
been determined without removing the target from the reaction chamber.

\subsection{Data Analysis}

The `LIST mode' approach of digital acquisition  is advantageous as it provides a sequential list of events with measured characteristics as needed by the user (energies and corresponding time information, as defined in the present experiment). By selecting a proper time window, the list mode data have been sorted. To analyze the list mode file and to  generate histograms
of energies within specific time windows, $MATLAB$ \cite{matlab} software package was used. 
For each group, a definite number of events (say,  20k) are sorted to create histogram that simply corresponds to the decay  spectrum for the desired time interval. The time resolution of the time-stamp is 10 ns. So, 20k events, in general, should correspond to a time interval of  at least 200 $\mu$s. The exact time difference is determined by taking the difference between the last and first time-stamps of each  set of 20k events. Thus, depending on the expected half-life of the radioactive nucleus, the time-interval for saving consecutive decay spectra to follow the decay can be suitably chosen.
 
In the present work (Fig. \ref{decay_IN}), event binning has been chosen such that reasonably good statistics remains even at the end of the decay. However, that reduces number of data points in the decay curve. On the other hand, smaller event binning yields larger number of data points in the decay curve, thus resulting in better fitting in some cases (Table \ref{half}). But the error in the area of each peak increases for smaller bin.  So, an optimization is needed depending on the source strength and half-life.  

For analysis of the decay spectra, INGASORT \cite{Ranjan} software  was used. The uncertainties shown in the Tables are arising from the errors in the fitting parameters. However, while fitting the decay curves as  shown in the figures, we have included the experimental errors (primarily statistical errors in the areas) in the data points.  No systematic error has been considered.

\section{Specific Experiments and Results}
\subsection{Experiment 1: Activation analysis }

The irradiation experiment has been performed at Variable Energy Cyclotron Centre, Kolkata using $\alpha$-beam from the K-130 cyclotron. A natural indium 
($^{115}$In (95.71\%), $^{113}$In (4.29\%)) \cite{Nuclear wallet card} target of thickness 54 $\mu$m is irradiated by $\alpha$-beam of 32 MeV energy. The $^{118}$Sb is the strongest populated nucleus in this experiment.
The ground state of $^{118}$Sb has a half-life of 3.6 (1) min, whereas one of its high spin ($J^\pi$ = 8$^-$) low energy isomeric state at 250 (6) keV excitation energy  has a half-life of 5.00 (2) h \cite{nndc} that decays via electron capture (EC) process to $^{118}$Sn. The irradiated foil has been carried  to a nearby laboratory for counting purpose after reduction of  activity to permissible limit to study the decay of this isomer. The data was collected by an 80\% relative efficiency HPGe detector along with CAEN 5780M Digitizer.

\begin{figure}[h]%[ht]
\begin{center}
\vspace{5cm}
\includegraphics[width=\linewidth]{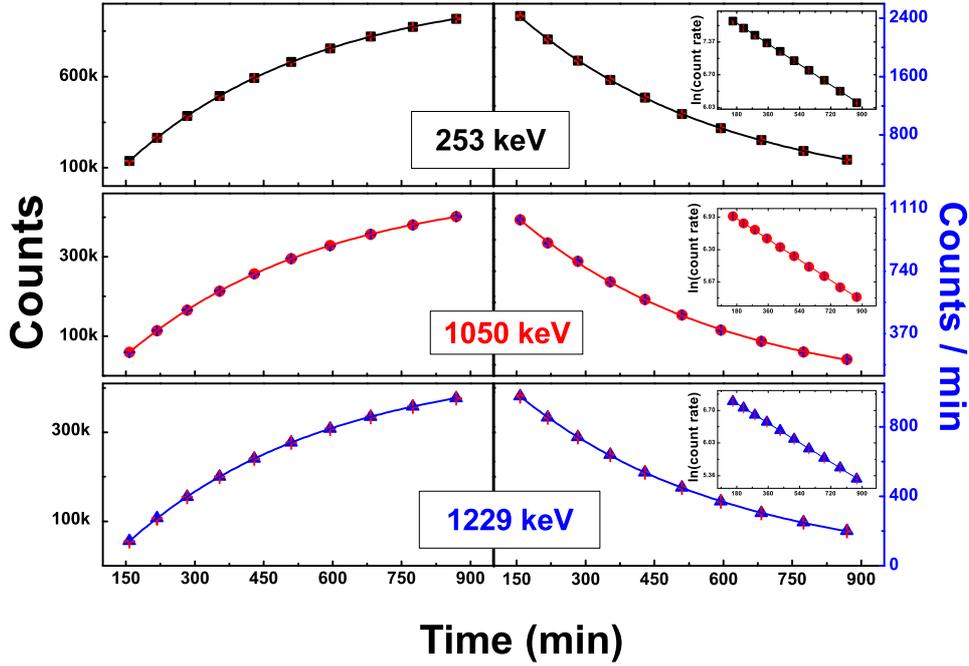}%{growth-decay.eps}%{decay_In_growtg.eps}
%\vspace*{-3cm}
\caption{\label{decay_IN} \small \sl (Color online) (Left:) Growth curves of  daughter nucleus ($^{118}$Sn) generated from yields of different 
decay $\gamma$-rays \cite{nndc} and (Right:) Decay of parent nucleus 
($^{118m}$Sb) observed   for different $\gamma$ transitions as function of time for event bin of 3M. The insets are decay curves plotted in log scale.}
\end{center}
\end{figure}

For  $^{118m}$Sb half-life measurement, the decay of parent nucleus ($^{118m}$Sb) and growth of daughter nucleus ($^{118}$Sn) have been plotted for different $\gamma$-rays emitted by the daughter nucleus (Fig. \ref{decay_IN}).  
For the decay curves, as time proceeds beyond one or more half-lives the statistics in the peaks of interest diminishes. The half-lives have been determined by fitting these curves with a
 single exponential decay function (Eqn. \ref{eqn4}). The $ln(Counts)$ are also plotted as a function of time  and a linear fitting has been done to extract the half-lives. 
Both the results are shown in the Table \ref{half}. The linear fitting gives better results as shown in the Table \ref{half}. We also have compared the results for two different event bins (3M and 1M events). Smaller event bin results in better fitting as more number of data points are present in the decay curve. However, the error in the area of each peak increases consequently. So, one has to optimise the event bin to have more accurate result.

For growth of daughter nuclei,  corresponding half-lives are determined by fitting the curves with the function mentioned in Eqn. \ref{eqn3}.
Table \ref{half} shows that lifetime measured by these two  different techniques are very similar. 
The present results are compared with the previous \cite{bolotin} results and they are found to  match well. The half-lives determined from growth curve show better agreement with previous  results.
As expected, the change in the sizes of the event bins does not show much variation in the results for growth curves.  Thus, it indicates that the time-stamped data of digital data acquisition system could be successfully used to evaluate the nuclear lifetime following the time evolution of the parent or daughter nucleus.

\begin{table}[ht]
\begin{center}
\caption{\label{half} Half-life of $^{118m}$Sb determined from time-evolution of decay $\gamma$s for two different event bins. The errors are indicated within brackets. }
\begin{tabular}{ccccccccc}

\hline 
 & \multispan{8} \hfil Half-life (h)\hfil\\ 
%\hrulefill \\
&\multispan{8}\hrulefill\\
Energy&  Prev.      & \multispan{7} \hfil Present \hfil\\
(keV)&\cite{nndc} &\multispan{7}\hrulefill\\
&& \multispan{4}\hfil Decay \hfil  && \multispan{2}\hfil Growth \hfil  \\
&&\multispan{4}\hrulefill&&\multispan{2}\hrulefill\\
&&  \multispan{2} \hfil Eqn. (\ref{eqn4}) \hfil&   \multispan{2} \hfil Linear \footnotemark[1] \hfil&&  \multispan{2} \hfil Eqn. (\ref{eqn3})  \hfil  \\
&&  BIN1\footnotemark[2]& BIN2\footnotemark[3]&  BIN1\footnotemark[2]& BIN2\footnotemark[3]&&  BIN1\footnotemark[2]& BIN2\footnotemark[3]    \\
\hline
253  &                   & 5.28 (5) & 5.25 (8)    & 4.95 (3) & 4.92 (2)& &4.87 (1)   & 4.87 (1)   \\\hline
1050 &    5.00 (2)       & 5.18 (15)& 5.07 (10)   & 5.20 (3) & 5.03 (2) && 5.00 (2) & 4.99 (1)    \\\hline
1229 &                   & 5.36 (7)& 5.23(10)    & 5.17 (2)& 5.11 (2)  && 5.04 (1) & 5.04 (1)         \\
\hline
\end{tabular}
\end{center}
\footnotemark[1]{Determined from logarithmic plot.  }\\
\footnotemark[2]{Event Bin of 3M.  }\\
\footnotemark[3]{Event Bin of 1M.  }\\

\end{table}

\subsection{Experiment 2: In-beam studies with CAEN digitizer }

In-beam experiment has been performed at  Variable Energy Cyclotron Centre, Kolkata, using  proton beam from the K-130 cyclotron. 
The natural Zn  target (Table \ref{decaymode}) was bombarded by 10 MeV proton beam, populating the compound nuclei, {\it viz.}, $^{65,67,68,69,71}$Ga isotopes, which de-excite by emitting  one  neutron and produce the residual nuclei 
$^{64,66,67,68,70}$Ga, respectively. The (p,n) reaction channel was the most dominant at this energy. The residual nuclei further undergo radioactive decay  as shown in Table \ref{decaymode}.
 
\begin{figure}[ht]
\begin{center}
\includegraphics[height=11.5cm, width=15 cm]{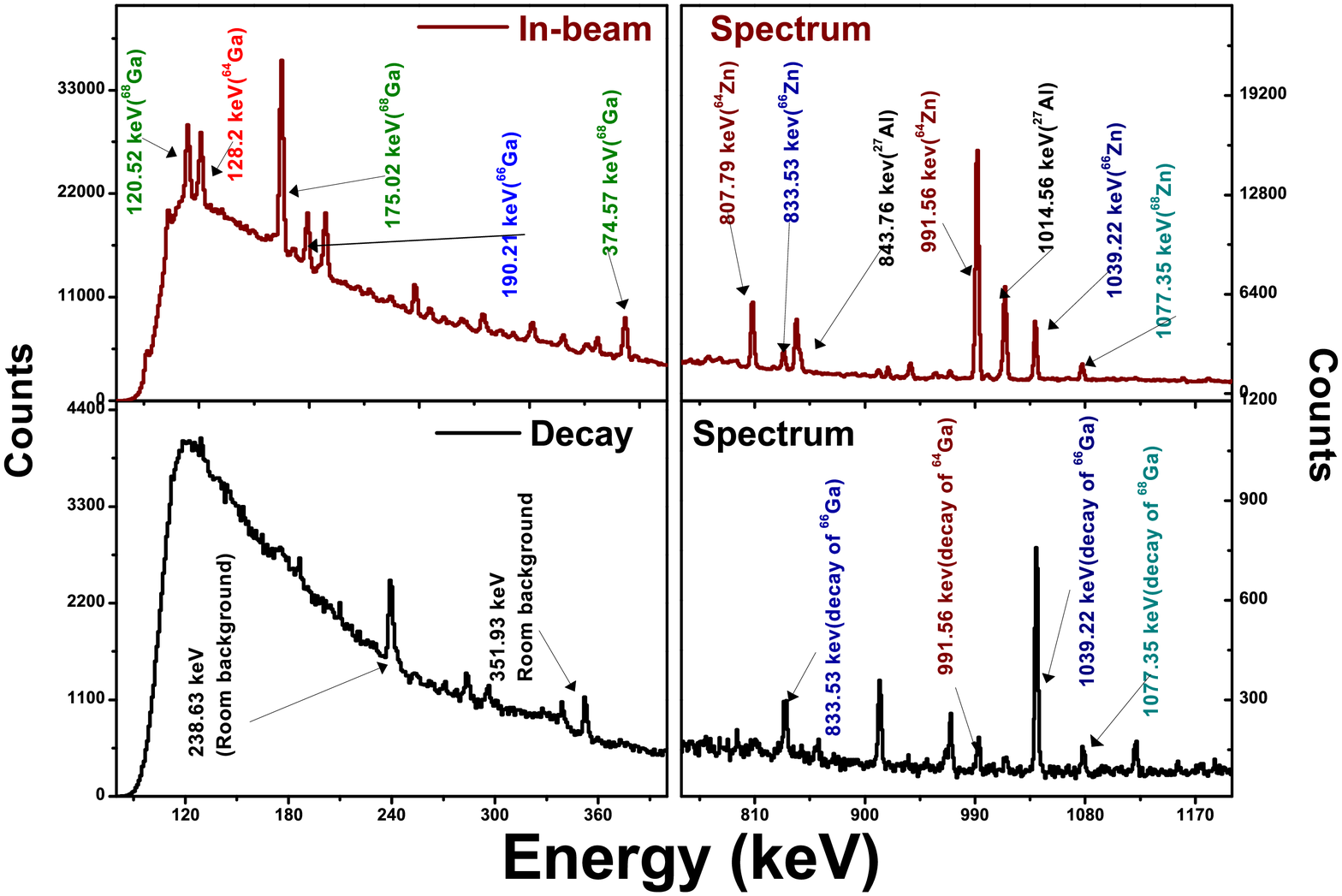}
\caption{\label{inbeam_zn} \small \sl (Color online) Typical in-beam and decay spectra of natural Zn target bombarded by 10 MeV proton projectile.  }
\end{center}
\end{figure}

 \begin{table}[ht]
\begin{center}
\caption{\label{decaymode} Isotopic composition of natural Zn and decay modes, half-lives and most intense decay $\gamma$-ray energies  of 
short-lived Ga isotopes produced by (p,n) reaction. The masses (shown in first column) are same for target Zn (Z=30) isotopes and corresponding Ga (Z=31) isotopes produced as reaction products. 
\cite{nndc}. }
\begin{tabular}{  c  c c c c c c }

\hline
Mass & Abundance &     Decay mode                  &  Half-life       &$\gamma$-ray Energy (keV)  \hfil\\
(A)&$^{nat}$Zn\cite{Nuclear wallet card} &  & (Error) &Intensity( \%)\footnotemark[1]\\

\hline
64   	&   49.17\%  &EC                         &2.627   min    & 991.51 \footnotemark[2]   \\
   	&        &                         &  (12)      &   (46)  \\
66      &   27.73\%   &EC                         & 9.49   h       & 1039.22 \footnotemark[2]  \\
    &         &                          &  (3)         &   (37) \\
67      &   4.04\%         &EC                         & 3.2617   d     & 93.31 \footnotemark[2] \\
      &              &                         &   (5)      &   (72.57)\\
68      &   18.45\%        &EC                         & 67.71   min    & 1077.34 \footnotemark[2] \\
      &             &                          &(8)      &  (3.22) \\
70      &   0.61\%         & $\beta^-$ (99.59\%)  &  21.14  min&1039.51\footnotemark[3] \\
     &          & EC (0.41\%)  & (5)&(0.65) \\
\hline
\end{tabular}
\end{center}
\footnotemark[1]{Absolute intensity per 100 decays.  }\\
\footnotemark[2]{From corresponding Zn isotopes via electron capture (EC)-decay }\\
\footnotemark[3]{From $^{70}$Ge via $\beta^-$-decay }

\end{table}
Thus, this experimental data contain information about the radioactive decay of  nuclei having half-lives ranging from minutes to days. The  $\gamma$-rays emitted by the  excited states of  nuclei 
produced in-beam as well as through decay (Fig. \ref{inbeam_zn}) are measured using a HPGe detector (with 20\% relative efficiency) in singles mode with CAEN 5780 digitizer as discussed before (Sec.2). 
In order to study the in-beam process, the time evolution of count has been determined for strong $\gamma$ transitions from 2$^{+}$ to 0$^{+}$ for $^{64,66,68}$Zn (991 keV, 1039 keV, 1077 keV) isotopes respectively.
These $\gamma$-rays are emitted from excited levels of $^{nat}$Zn isotopes populated by both $^{nat}$Zn(p,p$^\prime$)$^{64,66,68}$Zn reactions, as well as from the 
decay of $^{64,66,68}$Ga isotopes populated through $^{nat}$Zn(p,n)  reactions.   The 1014 keV $\gamma$ ray is emitted from the de-excitation of $^{27}$Al first excited state populated via (p,p$^\prime$) reaction. Aluminium is present in the target frame and other accessories in the chamber and thus can be considered as a measure of misalignment of the beam.

\begin{figure}[ht]
\begin{center}
\includegraphics[height=11.5cm, width=12.5 cm]{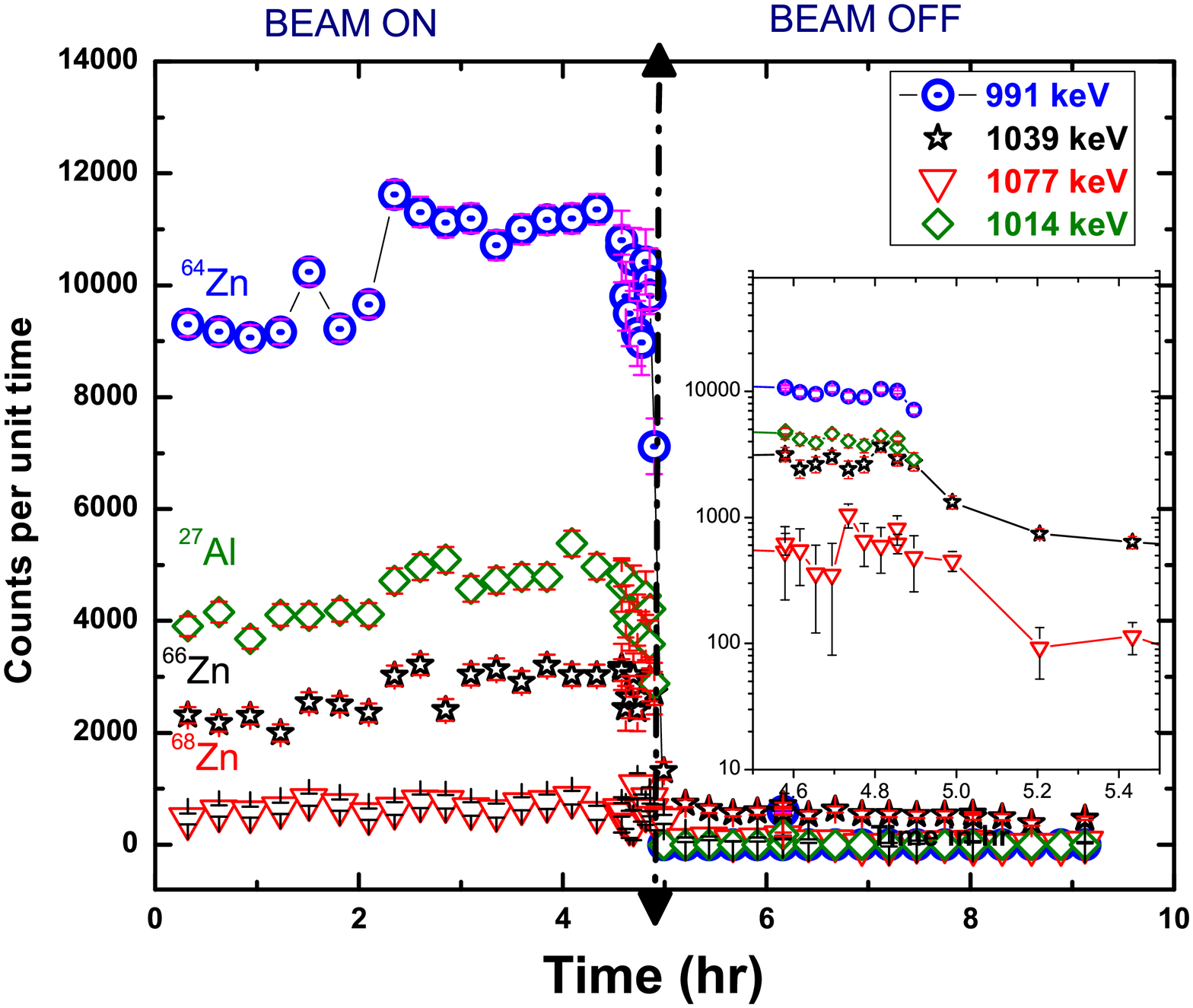} 
\caption{\label{znb2} \small \sl Variation of peak area of different $\gamma$-transitions with time. The dash-dotted line differentiates between the beam-on and beam-off situations. The inset shows a zoom-in of transition time. }
\end{center}
\end{figure}

The variations in counts for different $\gamma$ peaks have been plotted as a  function of time (shown in Fig.\ref{znb2}).  Four $\gamma$-ray energies E$_{\gamma}$ = 991, 1039, 1077 and 1014 keV correspond to the $\gamma$ transition of $^{64}$Zn, $^{66}$Zn, $^{68}$Zn and $^{27}$Al (target holder material) respectively. As can be seen from Fig.\ref{znb2}, the total counts under the $\gamma$-peaks remain almost the same during the BEAM-ON time. This is expected as there is no drastic change in beam parameters such as current intensity, beam focusing etc. \cite{DAE_2018}.

For a specific target material, in general, the absolute area under the photopeak is dependent on the abundance of the particular isotope, cross-section of the reaction as well as on beam parameters like beam current incident on the target. Moreover, the time evolution of the yield of  characteristic $\gamma$s emitted from the elements present in the target frame or the chamber can  also provide information regarding beam tuning and focusing. Such  details as a function of time can be obtained from these time-stamped data. Significant increase in the area of real peaks along with the beam-generated background peaks will be the evidence of an increase in beam current. Whereas, on the contrary, the increase in these background peaks compared  to the real peak, will be an indication of poor beam focusing.

In the BEAM-OFF condition, data acquisition continued for 4 hours. Since the half-life of $^{64}$Ga is of the order of a few minutes, it decays much faster  than $^{66,68}$Ga as can be seen  in Fig. \ref{znb2}. It is also noticed that the peak area under 1014 keV goes to zero  as an immediate effect of beam stop. The $^{27}$Al(p,n) reaction product $^{27}$Si has a half-life of 4.15 sec, which can also yield 1014 keV $\gamma$ ray.
However, the $^{27}$Si decays 99.7\% to the ground state of $^{27}$Al.   Thus, 1014 keV $\gamma$-ray is originating mostly from the in-beam excitation. 
From  Fig. \ref{znb2}, the yield curves of different Zn isotopes follow the  abundance ratio in $^{nat}$Zn. Their de-excitation $\gamma$s have nearly the same energy, which  ensures that the detector efficiency will be nearly the same.

\begin{table}[ht]
\begin{center}
\caption{\label{decay_ga} Measured  half-lives of short-lived Ga isotopes. }
\begin{tabular}{  c c c  c c }

\hline
Mass & Energy &\multispan{2} \hfil Half-life \hfil\\
&(keV)& \multispan{2}\hrulefill\\
&& Adopted \cite{nndc} & Present  work\\ \hline
64  &991 &2.627 $\pm$ 0.012 min  &2.61 $\pm$ 0.15 min  \\
\hline
66   &1039& 9.49 $\pm$ 0.03 h   & 9.82 $\pm$ 1.18 h \\
\hline
68   &1077& 67.71 $\pm$ 0.08 min & 70.30 $\pm$ 3.57 min  \\

\hline
\end{tabular}
\end{center}
\end{table}

\begin{figure}[ht]
\begin{center}
\includegraphics[height=11.5cm, width=12.5 cm]{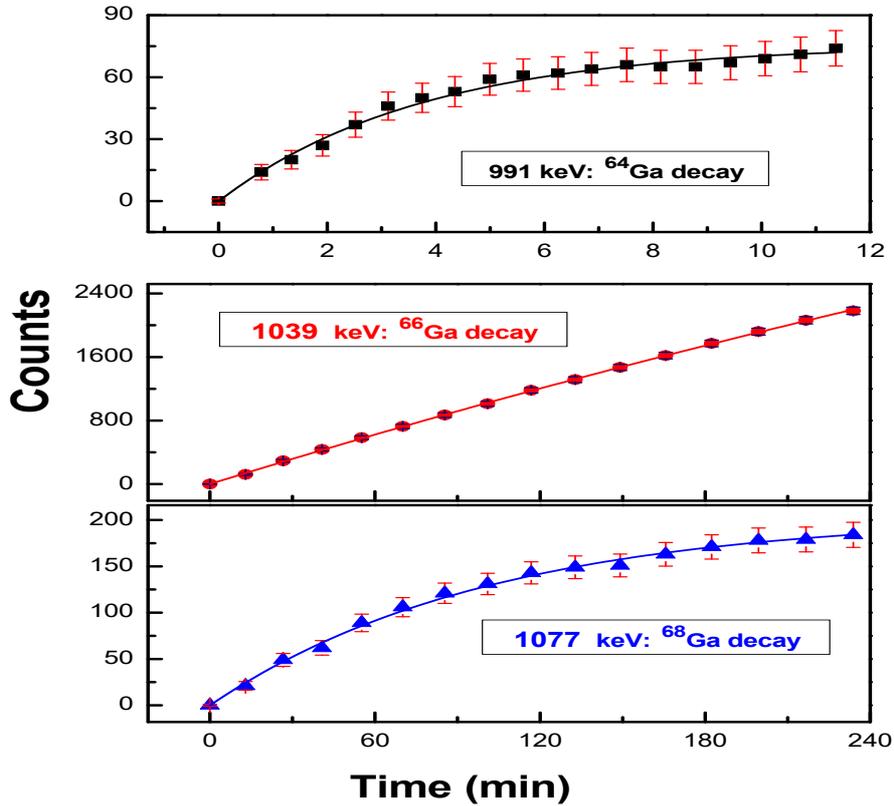}
\caption{\label{decay_zn_half} \small \sl (Color online) Growth curves of Zn isotopes ( $^{64,66,68}$Zn ) generated from decay gamma rays of corresponding radioactive Ga isotopes .}
\end{center}
\end{figure}

The half-lives of $^{64,66,68}$Ga isotopes have been extracted from the BEAM-OFF portion of the curve in Fig. \ref{znb2}. $^{67,70}$Ga decay data were not analysed due to long half-life ($^{67}$Ga) and weak decay $\gamma$-ray ($^{70}$Ga). In case of $^{64}$Ga, the growth  curve 
(Fig. \ref{decay_zn_half}) is plotted  for 12 minutes, beyond which it saturates. The growth curves (Fig. \ref{decay_zn_half}) of other two  daughter nuclei  ($^{66,68}$Ga), have been plotted up to $\sim$ 4 h  (depending upon the availability of data). The measured half-lives of $^{64}$Ga, $^{66}$Ga and $^{68}$Ga are tabulated in Table \ref{decay_ga}. The measured half-lives  match  reasonably well (Table \ref{decay_ga}) with the reported values \cite{nndc}. In the case of  $^{66}$Ga ($\tau_{1/2}\simeq 9.5$ h), large deviation of 
half-life from the reported value is due to the time restriction of our data  (only $\simeq 4 $ h data).  For $^{68}$Ga, relatively lower abundance of $^{68}$Zn (Table \ref{decaymode}) in natural Zn and lower intensity of the decay $\gamma$-ray increased the error in the result.   

\section{Discussions}

The time-stamped singles data recorded using the digital data acquisition system has been used to estimate the nuclear lifetime by measuring the $\gamma$ decay from an irradiated sample ($^{118}$Sb). The half-lives have been determined  from both the decay of parent nuclei and growth of daughter nuclei and  the results are found to be consistent with the previously reported value. With progression of time, the count rate decreases inducing larger statistical error in the decay. On the other hand, building up growth curve is advantageous as it reduces relative statistical error by resulting in better fit. Furthermore, this technique has been applied to estimate shorter half-lives for other nuclei ($^{64,66,68}$Ga) produced during the in-beam experiments in order to test the applicability. It is observed that the half-lives of $^{64,66,68}$Ga isotopes measured using the digitizer time-stamp match reasonably well with the values available in the literature. 

\section{Summary and conclusions}

We have determined  half-lives of  $^{64,66,68}$Ga isotopes and that of an isomeric state of $^{118m}$Sb nucleus, using digital time-stamped data acquired in singles mode. The half-lives of orders of minutes to days are determined from both the decay of parent  and growth of daughter nuclei, and found to be consistent with the previously reported values. This demonstrates the  effectiveness of time-stamped data for such measurements.  It would be very interesting to check  whether this technique is also  reliable for lifetime measurement in the range of a few tens of nanoseconds, which could not be tested in the present work.

\section{Acknowledgments}
The authors gratefully acknowledge Mr. Pradipta Kumar Das of Saha Institute of Nuclear Physics (SINP) and the target preparation facility of TIFR for preparation of the target. We want to thank Mr. Pradip Barua for his technical help. The authors acknowledge fruitful discussions with Dr. D. Banerjee of VECC during their irradiation experiment. The effort of all INGA collaborators are highly appreciated.

\end{document}